\documentclass[numerical,aip,reprint,apl]{revtex4-1}

\usepackage{amsmath,ulem}
\usepackage{graphicx}
\usepackage{dcolumn} 

\begin{document}

 \preprint{APS/123-QED}
 
 \title{Ge condensation under SiGe oxidization: from Molecular
   Dynamics simulation to one-dimensional analytic modeling. }

\author{Patrick Ganster} 
\affiliation{CINaM, UMR 7325 CNRS, Aix-Marseille University, 
Campus de Luminy, 13288 Marseille cedex 9, France}
\email[email: ]{ganster@emse.fr}
\altaffiliation[Present address: ]{MSE, SMS-MPI, 158 cours Fauriel, 42023 Saint-Etienne Cedex 2, France}

\author{Andr\`es Sa\'ul}
\email[email: ]{saul@cinam.univ-mrs.fr}
\affiliation{CINaM, UMR 7325 CNRS, Aix-Marseille University, 
Campus de Luminy, 13288 Marseille cedex 9, France}

\author{Guy Tr\'eglia} 
\email[email: ]{treglia@cinam.univ-mrs.fr}
\affiliation{CINaM, UMR 7325 CNRS, Aix-Marseille University, 
Campus de Luminy, 13288 Marseille cedex 9, France}
\email{treglia@cinam.univ-mrs.fr}

\date{\today}

\begin{abstract}
  Oxidization of a dilute Si(Ge) alloy is modeled using an original
  protocol based on molecular dynamics simulation and rules for the
  oxygen insertions. These rules, deduced from {\it ab-initio}
  calculations, favor the formation of SiO$_2$ against GeO$_2$ oxide
  which leads to segregation of Ge atoms into the alloy during the
  oxidization front advance. Ge condensation is then observed close to
  the SiO$_2$/Ge interface due to the strain induced by oxydization in
  this region. From the analysis of the simulation process, we propose
  a one-dimensional description of Ge condensation wich perfectly
  reproduces the evolution of the Ge concentration during oxidization
  of the SiGe alloy.
\end{abstract}

\pacs{Valid PACS appear here}
\maketitle
  
In the challenging world of nanotechnologies, the so-called,
Ge-condensation technique, is a promising way to realize thin Ge
layers of good quality for high mobility MOSFET channel
\cite{Dissanayake2008178}. The method consists in oxidizing at high
temperature an Si$_{1-x}$Ge$_x$ alloy epitaxially grown on SOI
substrate and preliminary capped by a thin silicon layer. After the
oxidization of the silicon cap that forms SiO$_2$ oxide, the pursuit
of oxidization induces the segregation of Ge atoms in the SiGe alloy
which becomes enriched in Ge.

The development of predicting numerical tools to optimize
manufacturing steps \cite{TCAD} is essential and necessary to get a
good understanding of the relevant mechanisms involved during the
Ge-condensation process. In the context of the continuous decrease of
the microsystem sizes and the use of smaller quantities of matter,
descriptions and precise characterizations at the atomic scale is a
necessary road map to progress in the understanding of the driving
mechanisms involved in the Ge condensation technique. Such an
atomistic description is a challenging problem which needs to deal
with a complex system and many mechanisms occurring at the same time.
In particular, it requires to model the oxidization of the SiGe alloy
and the segregation of Ge atoms from the oxydised region to the
substrate.  A full description taking into account all mechanisms and
reproducing the time evolution of the system in one model is obviously
far from the actual state-of-art of computational science at atomic
scales.

In this letter, we propose an atomistic model which couples molecular
dynamics simulations and rules deduced from {\it ab-initio}
calculations to model oxidization of SiGe alloy involving the Ge
segregation.  In order to be able to describe a system sufficiently
large to be representative of the Ge condensation, we use the
Stillinger-Weber (SW) empirical potential which has been successfully
employed and optimized to describe SiGe alloys and Si-O mixed systems
by Molecular Dynamics or Monte Carlo simulation
\cite{Krishnan20072975, Tzoumanekas2000670, PhysRevB.53.8386, clancy,
  watanabe, DallaTorre, ganster}.  The parameters of this potential,
which is composed of 2-body and 3-body interactions, were fitted to
{\it ab-initio} calculations and thermodynamical properties of Si, Ge,
and SiGe compounds.  For the mixed Si-Ge-O system, we derived a set of
parameters from those developed by Watanabe {\it et al.}
\cite{watanabe} and Clancy {\sl et al.} \cite{clancy} to study
respectively Si-O and a SiGe. In practice, for the Ge-O interactions,
we scaled the SW parametrization of Watanabe {\sl et al.}  by using
the size and energy parameters of the Ge parametrization developed by
Clancy {\sl et al.} \cite{clancy}.  Arithmetical and geometrical
averages have been used for the size and energy parameters
respectively the in 3-body interactions involving triplets composed of
Si, O and Ge atoms such as OSiGe, SiSiGe, SiOGe, ...

As mentioned above, the Ge condensation process begins by the
oxidization of the silicon cap covering the SiGe alloy.  This step is
essential since it initiates the formation of a SiO$_2$ oxide
\cite{Di2005}. After the oxidation of the whole silicon cap, one would
expect the formation of either SiO$_2$, GeO$_2$ or a mixed
SiO$_2$-GeO$_2$ amorphous oxide. However, the experiments show a
preferential oxidization of the Si atoms which induces the segregation
of Ge atoms and an increase of the fraction of Ge in the SiGe alloy
\cite{Dissanayake2008178, Li2003127}.  This preferential oxydization,
can be understood from the differential in formation energy $E_f$
between SiO$_2$ and GeO$_2$ in the $\alpha$-quartz phases :
\begin{eqnarray} \label{ef}
 \text{Si} + \text{O}_2 \longrightarrow \text{SiO}_2 \nonumber \\
 \text{Ge} + \text{O}_2 \longrightarrow \text{GeO}_2 \nonumber
\end{eqnarray}
From the total energies obtained using the \textsc{Wien2k} code
\footnote{The \textsc{Wien2k} program is an implementation of the
  full-potential linearized augmented plane-wave method based on
  density-functional theory. P. Blaha, K. Schwarz, G.  Madsen, D.
  Kvaniscka, and J.  Luitz, in Wien2k, An Augmented Plane Wave Plus
  Local Orbitals Program for Calculating Crystal Properties, edited by
  K. Schwarz Vienna University of Technology, Vienna, Austria, 2001.}
the formation energy can be calculated as :
\begin{equation}
E_f^{\text{XO}_2} = E^{\text{XO}_2} - E^{\text{X}} -
E^{\text{O}_2}
\end{equation}
where $E^{\text{XO}_2}$ is the total energy of the oxide,
$E^{\text{X}}$ is the energy of $X$ = Si or Ge in the diamond phase
structure and $E^{\text{O}_2}$ the energy of an isolated $O_2$
molecule. We have found that the $E_f^{\text{SiO}_2}$ is much lower
than $E_f^{\text{GeO}_2}$, respectivelly -8.2 eV and -4.7 eV
\footnote{The calculations presented here were performed using the
  generalized gradient approximation of Perdew, Burke, and
  Ernzerhof~\cite{Perdew96} (GGA) for exchange and correlation and a
  cut-off parameter $RK_\text{max}=8$.}, in full qualitative agreement
with experimental observation since $\Delta $ = $E_f^{GeO_2}$ -
$E_f^{SiO_2}$ = 3.5 eV implies that the formation of $SiO_2$ is highly
favored against that of $GeO_2$.

The vality of our parametrization of the semi-empirical SW potential
was also checked on its ability to describe this preferential
oxydation.  The difference between the formation energy of SiO$_2$ and
GeO$_2$ was found to be $\Delta = 1.8$ eV, which is half the value
obtained by the {\it ab-initio} calculations, but widely sufficient to
ensure the empirical model to clearly favor the formation of SiO$_2$
against GeO$_2$ phase.

\begin{figure}[htb]
\includegraphics[width=8cm]{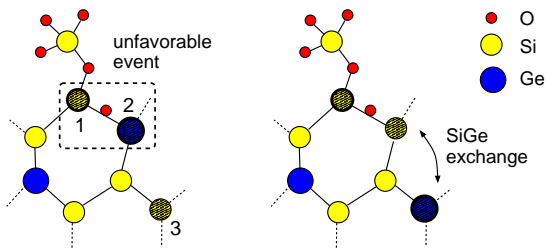}
\caption{\label{fig:scheme} 
  (color online) Scheme of the oxygen insertion taking into account
  the preferably oxidization of the Si atom i) the highest Si in the
  substrate surrounded by at least one and at most three oxygen atoms
  is located (yellow circle labeled 1), ii) when the last chance to
  introduce oxygen is on a Si-Ge bond, the considered Ge atom (bleu
  circle numebered 2) is exchanged with a non-oxidized Si atom of the
  substrate (yellow circle labeled 3). }
\end{figure}

The simulation box that we have used was oriented in such a way that
the $x$-, $y$-, and $z$-axis of the simulation box are along the
$<$100$>$ directions of the diamond lattice. The system size is 27.15
x 27.15 x 108.6 \AA$^3$.  Periodic boundary conditions were applied in
the [100] and [010] directions (\textit{i.e.} $x$ and $y$) and the
system was free to relax in the [001] direction (free surfaces).  The
starting configuration was a bare Si$_{0.95}$Ge$_{0.05}$ substrate
(3700 Si and 194 Ge atoms), which was capped by a pure Si layer in the
upper part (300 Si additional atoms).

To simulate SiGe oxydization, we use a protocol based on molecular
dynamics (MD) simulations and on the one-by-one incorporation of
oxygen atoms into Si-Si bonds. This protocole is an extension to the
one that we have previously used in the case of Si oxydization
\cite{ganster}. We take now into account the preferential formation of
SiO$_2$ against $GeO_2$ oxide. In order to simulate the initial nuclei
of the oxidization process, oxygen atoms are placed on 10~\% of the Si
dangling bonds (randomly chosen) on top of the cap Si layer, which is
not reconstructed before the oxygen insertion. This initial
configuration is equilibrated at 1200~K during 20~ps by rescaling the
atomic velocities every 0.5~ps.  Then, we include the oxygen atoms one
by one following the following sequence. We search for the highest Si
atom in the substrate which is bound to at least one and at most three
oxygen atoms and the oxygen atom is placed in the middle of the
longest Si-Si bond available around this Si atom \cite{ganster}. When
only a Si-Ge bond is found, we place the oxygen atom in the middle of
the bond and we exchange the Ge atom with a non-oxidized Si atom of
the SiGe substrate. The latter being determined in order to be the
most energetically favorable exchange.  This procedure is schematized
in the Fig.\ \ref{fig:scheme}.  The whole system is relaxed by a
Molecular Dynamic run at 1200~K during 1 ps. The full scheme
(insertion, relaxation) is repeated for 6000 oxygen atoms insertions
during which some 600-700 Si-Ge exchanges were realized.  This
procedure leads to the formation of a pure amorphous SiO$_2$ oxide
which is composed of a random network of SiO$_4$ entities connected by
vertex, which is very similar to that of the SiO$_2$/Si system
\cite{ganster}.

\begin{figure}[b]
\includegraphics[width=7cm]{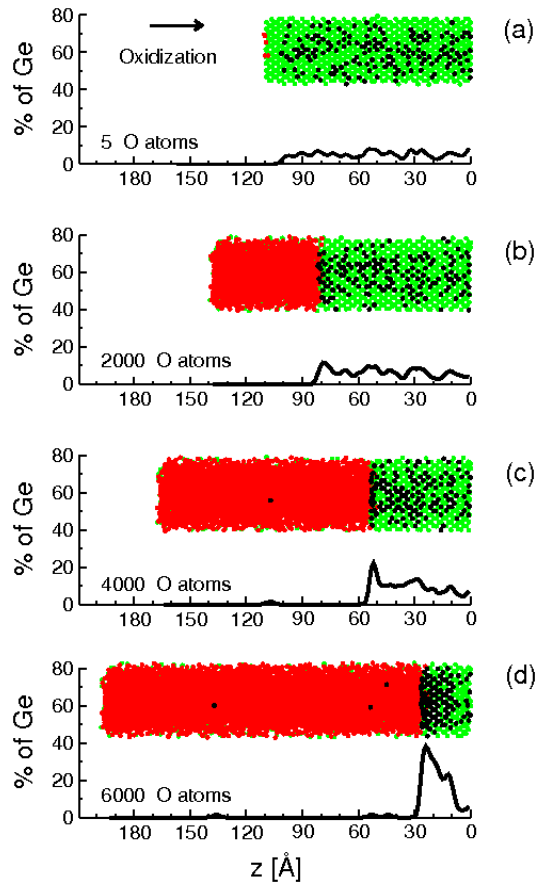} 
\caption{\label{fig:concentration} (color online) 
  Evolution of the Ge concentration
  $n_\text{Ge}/(n_\text{Ge}+n_\text{Si})$ profile for the SiO$_2$/SiGe
  system after insertion of 5, 2000, 4000, and 6000 oxygen atoms. O,
  Si and Ge atoms are represented in red, green and black
  respectively. }
\end{figure}

The Ge concentration profile along the $z$-axis below the SiO$_2$/SiGe
interface for a typical run is shown in Fig.\ \ref{fig:concentration}
after different oxidation steps (5, 2000, 4000, and 6000 oxygen
insertions).  The concentration is defined for diffuse layers of
thickness $\Delta z$ = 1 \AA.  Starting from an uniform concentration
of Ge along the $z$-axis (Fig.  \ref{fig:concentration}(a)), one can
see that the Ge atoms concentrate in the vicinity of the interface
between SiO$_2$ and SiGe (Fig.  \ref{fig:concentration}(b)-(d)), in
good agreement with RBS measurements \cite{legoues:644}.

The overal picture is that the oxidation evolves by pushing a Ge
strip which thickens all along the process, whilst preserving a
concentration lower than 50 \% per plane.  This means that the Ge
atoms expelled from the oxide mainly exchange with Si substrate
atoms located close to the SiO$_2$/SiGe interface, even though a few
exchanges also involve Si atoms far from this interface, leading to a
slight increase of the overall Ge concentration.

\begin{figure}[h]
\includegraphics[width=8cm]{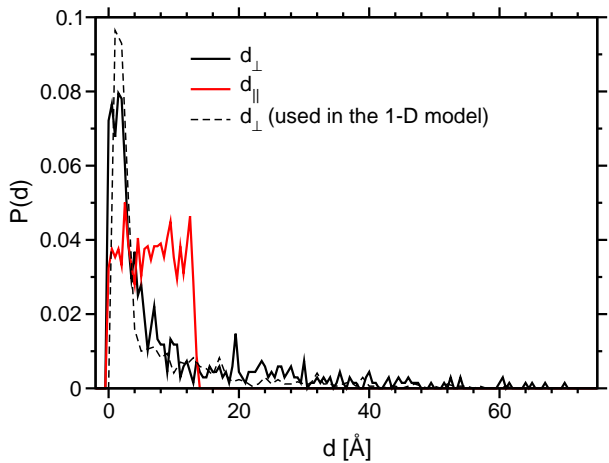}
 \caption{\label{fig:exchange} 
   (color online) Distribution of the parallel $d_{\parallel}$ and
   perpendicular $d_{\perp}$ projections of the distances separating
   exchanged Si and Ge atoms during the oxidation process. }
\end{figure}

This qualitative behaviour can be confirmed by analyzing the relative
position of the Si and Ge atoms which are exchanged during the
oxidation process.  In Fig.\ \ref{fig:exchange} we show an histogram
of the projections of the distance parallel $d_{\parallel}$ and
perpendicular $d_{\perp}$ to the interface. The distribution of
$d_{\parallel}$ is uniform and does not exceed half the size of the
simulation box, which indicates that exchanges occur between Si and Ge
atoms ranging for all the simulation box width.  On the contrary, the
distribution of $d_{\perp}$ presents a narrow peak which is almost
completely damped beyond 5 \AA\ which corresponds to three (001)
interatomic layers.  This shows that the Ge atoms mainly exchange with
Si atoms located in the vicinity of the SiO$_2$/SiGe interface, which
leads to their accumulation (condensation) in this limited region.
The tail of the distribution of $d_{\perp}$ towards larger distances
confirms that a few exchanges can also occur between more distant Si
and Ge atoms, leading to the slight increase in the overal Ge
concentration profile.

\begin{figure}[htp]
\includegraphics[width=9cm]{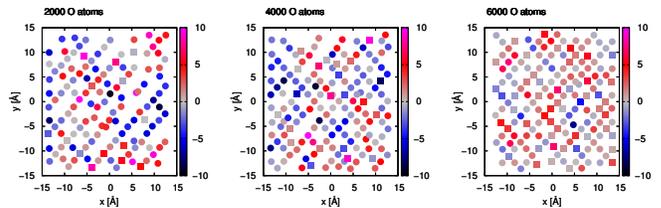}
 \caption{\label{fig:strain} 
   (color online) Lateral stress maps of the first three layers below
   the SiO$_2$/SiGe interface after insertion of 2000, 4000 and 6000
   oxygen atoms. Tensile and compressive regions are represnted in
   blue and red respectivelly.}
\end{figure}

It remains to identify the physical origin of the preference for Si-Ge
exchanges to occur close to the interface.  
A carefull analysis of the local concentrations does not allow to
establish any correlation between the Ge concentration in the SiGe
substrate and the new local environment of the exchanged Ge atom.
%
However, we have found that the this preference is induced by the
strain originated by the oxidation in the interfacial region.  In
Fig.\ \ref{fig:strain} we show the pressure maps averaged over the
first three SiGe layers below the oxidation front after insertion of
2000, 4000 and 6000 oxygen atoms.
These two dimensional maps, parallel to the interface, are very
similar at each step of the process and reveal an inhomogeneous strain
distribution which is rapidly damped beyond these three layers.  More
precisely, one sees that tensile and compressive regions coexist in
almost equal proportions.  The Ge condensation close to the interface
is then originated by the segregation of the "big" Ge atoms (the
size-mismatch between Si and Ge is of about 4 \%) in the tensile
regions. The number of tensile sites decreases when the Ge
concentration increases and should disappear when the concentration
reaches 50\%.  This is indeed what is observed in Fig.\ 
\ref{fig:concentration}(d), confirming the strain origin of Ge
condensation close to SiO$_2$/SiGe interface.

From the previous analysis, it is clear that one of the essential
ingredients which are responsible of the Ge condensation, is the
peculiar distribution of distances between exchanged Si-Ge atoms
displayed in Fig.\ \ref{fig:exchange}).  A simple one dimensional
model of Ge condensation can be used to understand the role played by
the overall shape of this distribution. Let us consider a non-oxidized
SiGe substrate composed of $N$ layers labelled $p$ ($p = 1, N$), each
layer contains a finite number of atoms $n(p) = n_\text{Ge}(p) +
n_\text{Si}(p)$ where $n_\text{Ge}(p)$ and $n_\text{Si}(p)$ are
respectively the number of Ge and Si atoms in the $p$ layer. Assuming
that oxidation is progressing layer-by-layers, by exchanging Ge atoms
from the first $p$ layers, starting from $p=1$, with Si atoms of
deeper layers, distant by $d_\perp$ from the former, the Ge
condensation follows the simple set of equations :

\begin{eqnarray} 
   n_{Ge}(p)   & = & n_{Ge}(p) - 1   \label{geminus}      \\
   n_{Si}(p)   & = & n_{Si}(p) + 1   \label{siplus}      \\ 
   n_{Ge}(p+d_\perp) & = & n_{Ge}(p+d_\perp) + 1 \label{geplus}      \\
   n_{Si}(p+d_\perp) & = & n_{Si}(p+d_\perp) - 1 \label{siminus}      
\end{eqnarray} 

\begin{figure}[htb]
\begin{center}
  \includegraphics[width=8cm]{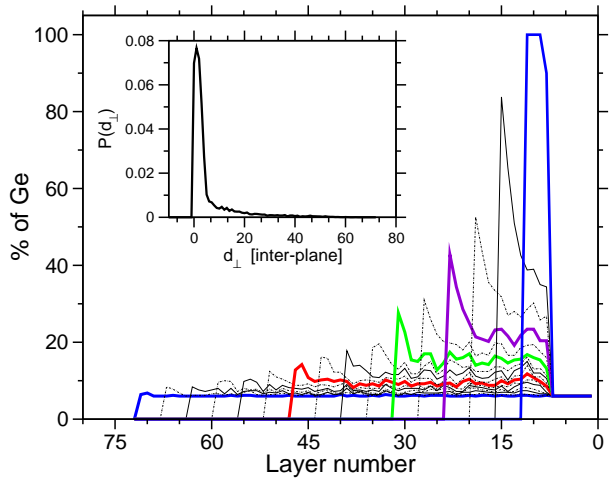}
  \caption{\label{fig:concentration-1D} Ge concentration after 
    each four layers depleted in Ge.  The inset shows the distribution
    of SiGe exchanged during the simulation.}
\end{center}
\end{figure}

Starting from the same initial Ge concentration 5 \%, number of atoms
per layer $n(p) = 50$ and number of SiGe layers $N=72$ than in the MD
simulation, the Equations (\ref{geminus})-(\ref{siminus}) lead to the
evolution of the Ge concentration profiles displayed in Fig.\ 
\ref{fig:concentration-1D}.  The inset in this figure recalls the
probability distribution of the interlayer distance $d_\perp$ between
the exchanged Si and Ge atoms which is used in the simulation, i.e.,
for each exchange we randomly choose a $d_\perp$ following this
distribution.

As can be seen, this simple model reproduces the essential features of
the MD model, namely the formation of a Ge strip below the
SiO$_2$/SiGe interface, which thickens as oxidation goes along.
However, the model misses the saturation of Ge segregation at about 50
\% per plane.

To summarize, we have shown here by using atomistic models combining
{\it ab-initio} and empirical potentials), that the main driving force
for Ge condensation under oxidation are the preferential SiO$_2$
formation against that of GeO$_2$ and the existence of an
inhomogeneous strain extending on a few layers below this interface.
The latter effect is at the origin of the localisation of the Ge
agregates close to the interface. In addition, we have shown that a
simple 1D model, based on the single knowledge of the distribution of
distance between Ge and Si atoms exchanged during the simulation
(related to the strain distribution), is sufficient to account for the
main features of Ge condensation, providing us a flexible method to
study this phenomenon under various conditions.
  
\bibliographystyle{apsrev4-1}

\bibliography{papersige}

\end{document}